\documentclass[multphys,vecphys]{svmult}
\usepackage{amsfonts}
\usepackage{amssymb}

\usepackage{makeidx}         
\usepackage{graphicx}        
\usepackage{multicol}        
\usepackage[bottom]{footmisc}

\makeindex             
\usepackage{url}

\DeclareMathSymbol{\la}{3}{AMSa}{46}
\DeclareMathSymbol{\ga}{3}{AMSa}{38}

\begin{document}


\title*{Power Spectrum Estimation II. Linear Maximum Likelihood}
\author{Andrew J S Hamilton}
\institute{JILA and Dept.\ Astrophysical \& Planetary Sciences,
Box 440, U. Colorado, Boulder, CO 80309, USA.
\texttt{Andrew.Hamilton@colorado.edu \url{http://casa.colorado.edu/~ajsh/}}}




\newcommand{\dd}{\D}
\newcommand{\ddd}{\dd^3}
\newcommand{\e}{\E}
\newcommand{\im}{\I}
\newcommand{\transpose}{\top}
\newcommand{\qex}{\nopagebreak[4]$\qed$\pagebreak[2]}

\newcommand{\el}{\ell}
\newcommand{\nbar}{{\bar n}}
\newcommand{\deltatilde}{{\tilde\delta}}
\newcommand{\ntilde}{{\widetilde n}}
\newcommand{\Ptilde}{{\widetilde P}}

\newcommand{\kvec}{{\vec{k}}}
\newcommand{\Lvec}{{\vec{L}}}
\newcommand{\nvec}{{\vec{n}}}
\newcommand{\pvec}{{\vec{p}}}
\newcommand{\qvec}{{\vec{q}}}
\newcommand{\rvec}{{\vec{r}}}
\newcommand{\vvec}{{\vec{v}}}
\newcommand{\xvec}{{\vec{x}}}
\newcommand{\yvec}{{\vec{y}}}
\newcommand{\zvec}{{\vec{z}}}
\newcommand{\onevec}{{\vec{1}}}

\newcommand{\khat}{{\hat\kvec}}
\newcommand{\rhat}{{\hat\rvec}}
\newcommand{\zhat}{{\hat\zvec}}

\newcommand{\mangle}{{\sc mangle}}

\newcommand{\skipp}{\vskip.25cm\noindent{}}

\newcommand{\xikfig}{
    \begin{figure}[t]
    \begin{center}
    \includegraphics[scale=.6]{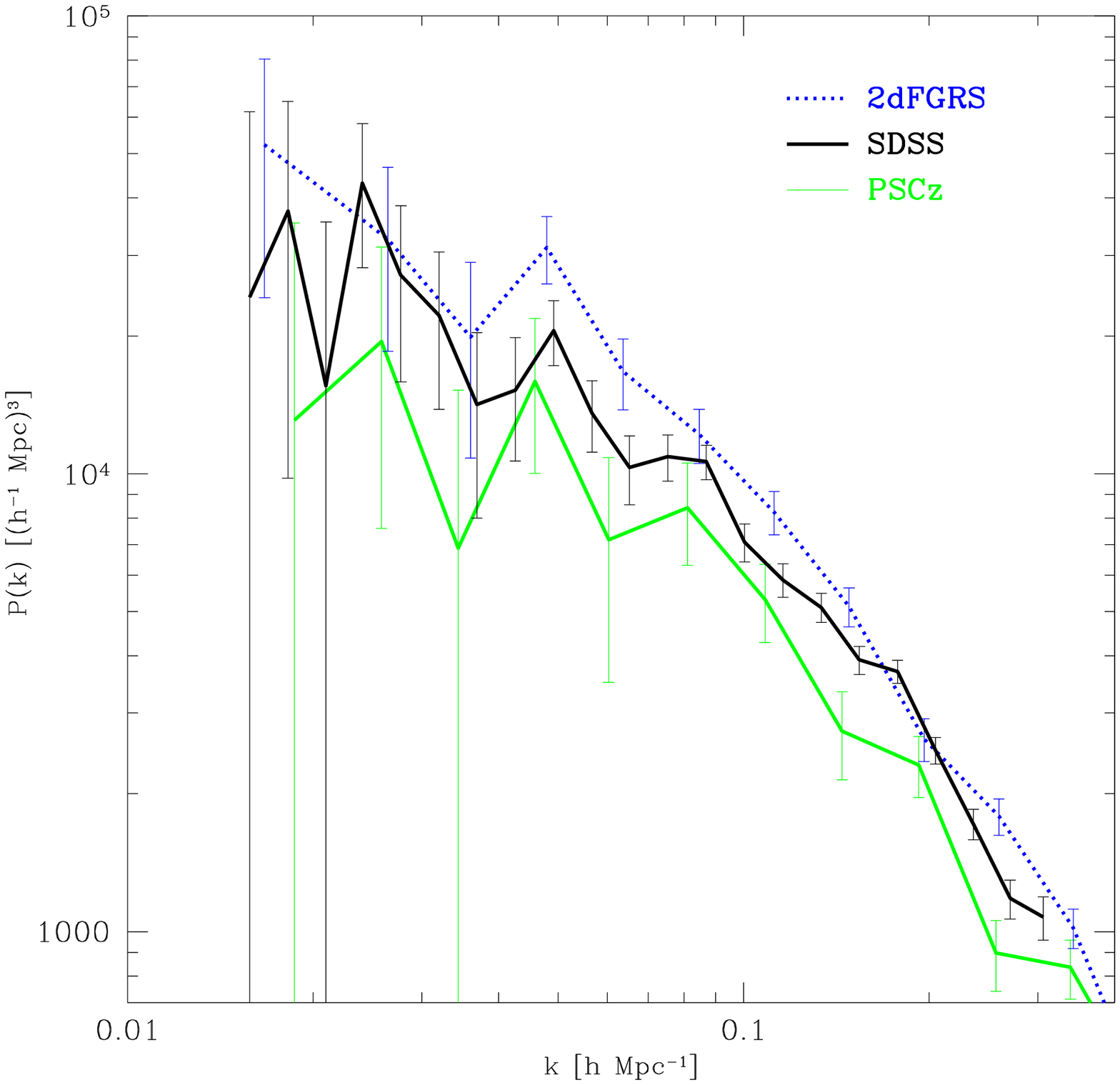}
    \end{center}
    \caption[1]{
    \label{xik}
From Tegmark et al (2004) \cite{T04}.
Decorrelated, real space (not redshift space) linear power spectra
measured from the PSCz survey
(Hamilton, Tegmark \& Padmanabhan 2000 \protect\cite{HTP00}),
the 2dF 100k survey
(Tegmark, Hamilton \& Xu 2002 \protect\cite{THX02}),
and the SDSS survey
(Tegmark et al 2004 \protect\cite{T04}).
    }
    \end{figure}
}

\newcommand{\mappsczfig}{
    \begin{figure}[p!]		
    \begin{center}
    \includegraphics[scale=.36]{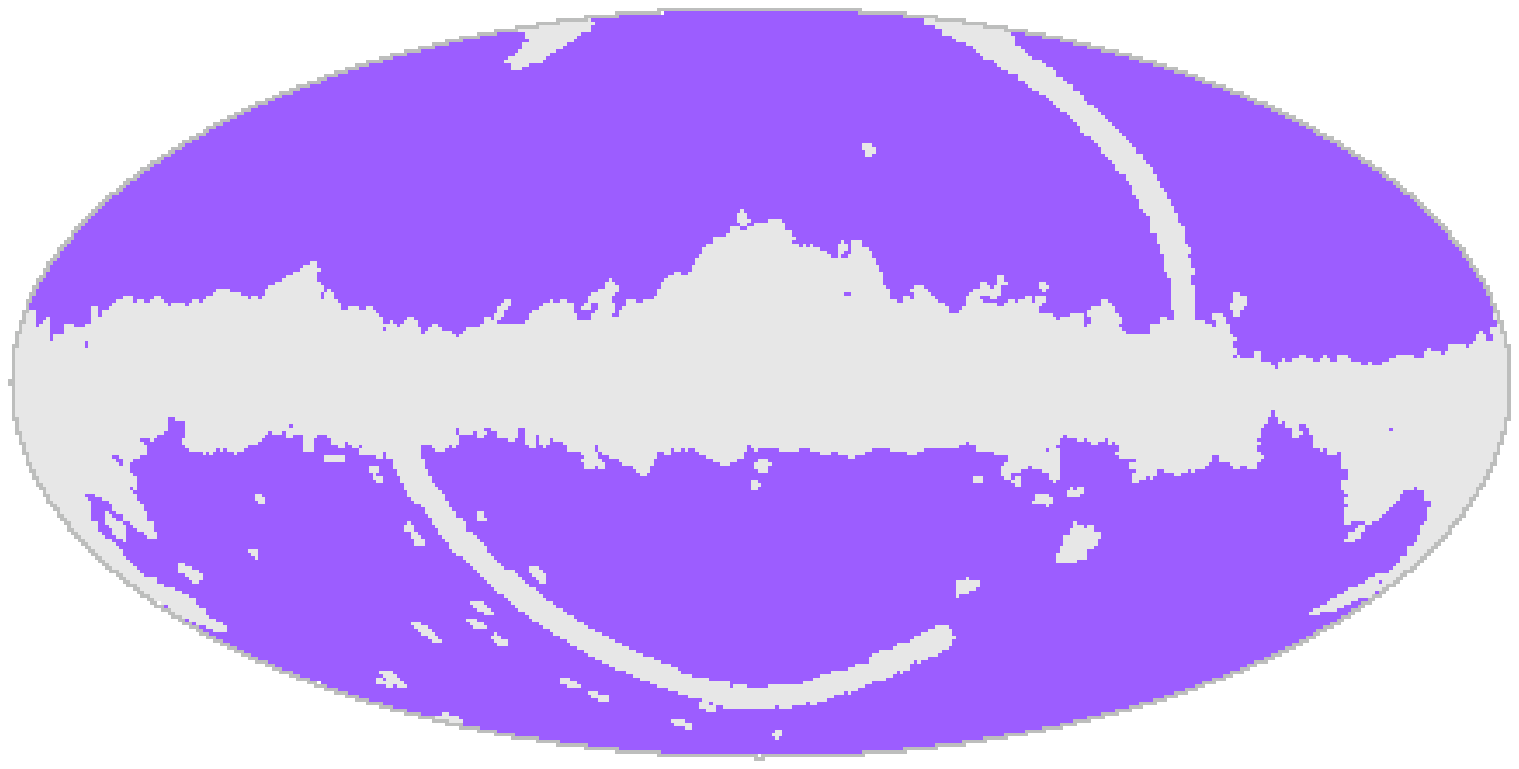}
    \includegraphics[scale=.36]{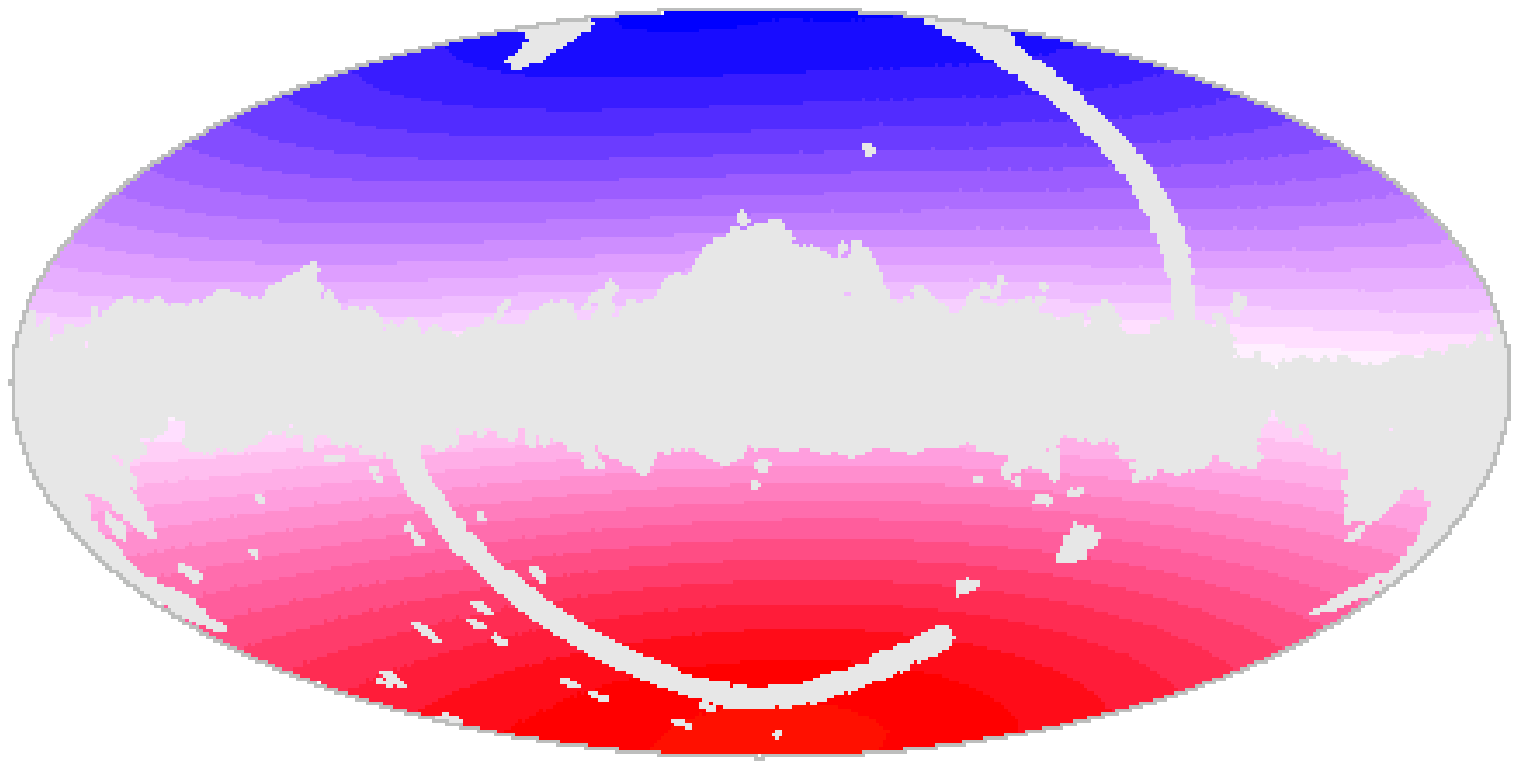}
    \includegraphics[scale=.36]{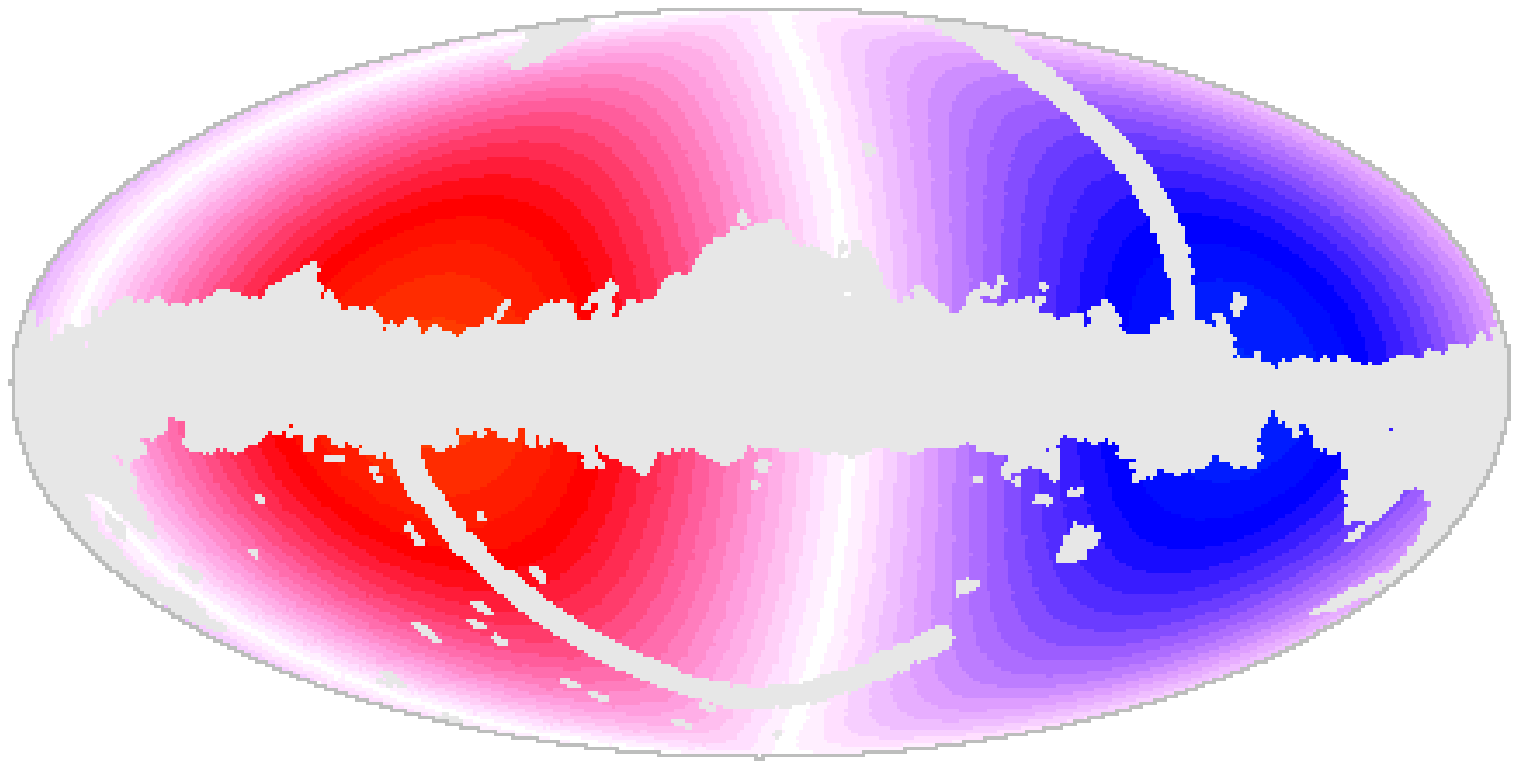}
    \includegraphics[scale=.36]{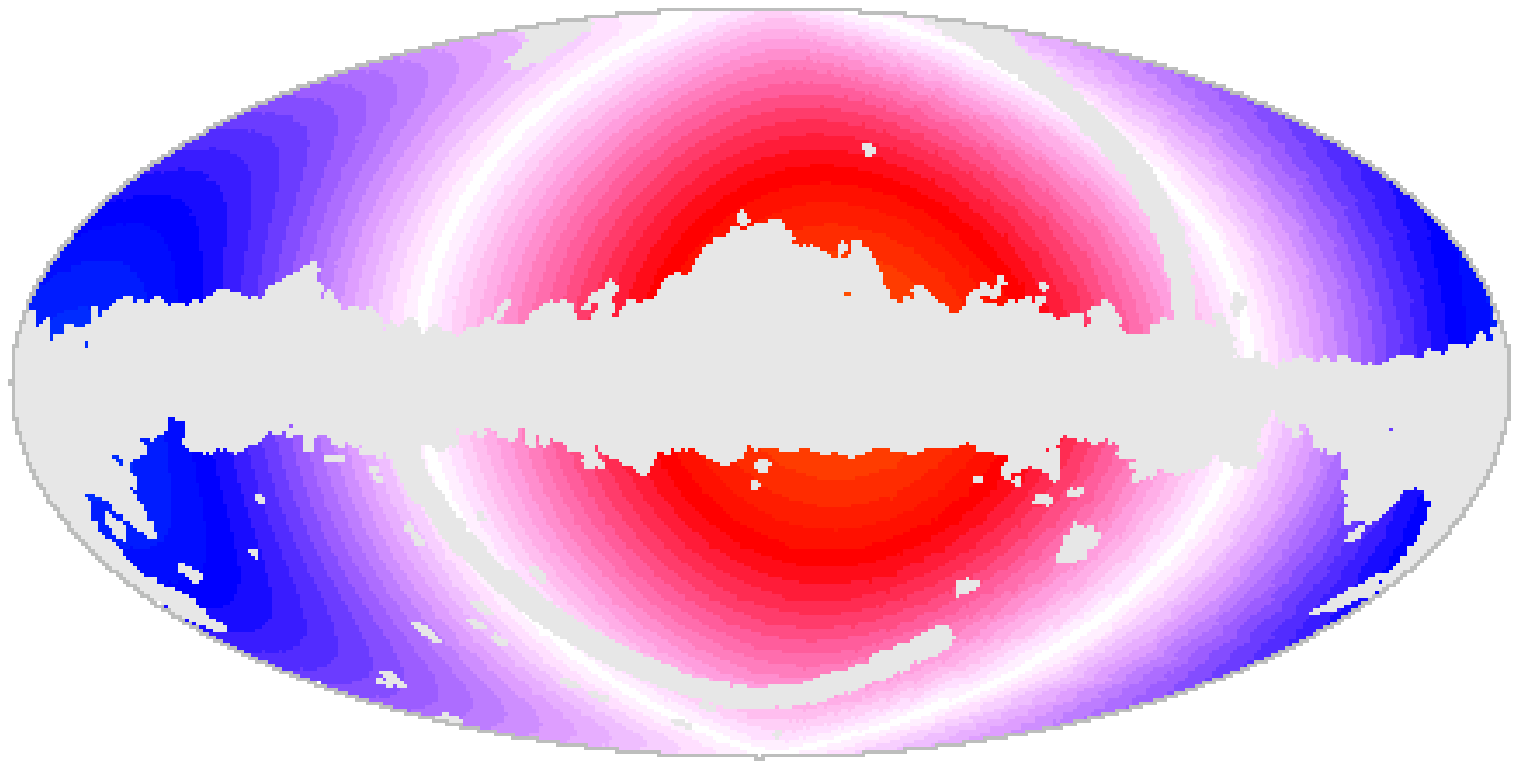}
    \includegraphics[scale=.36]{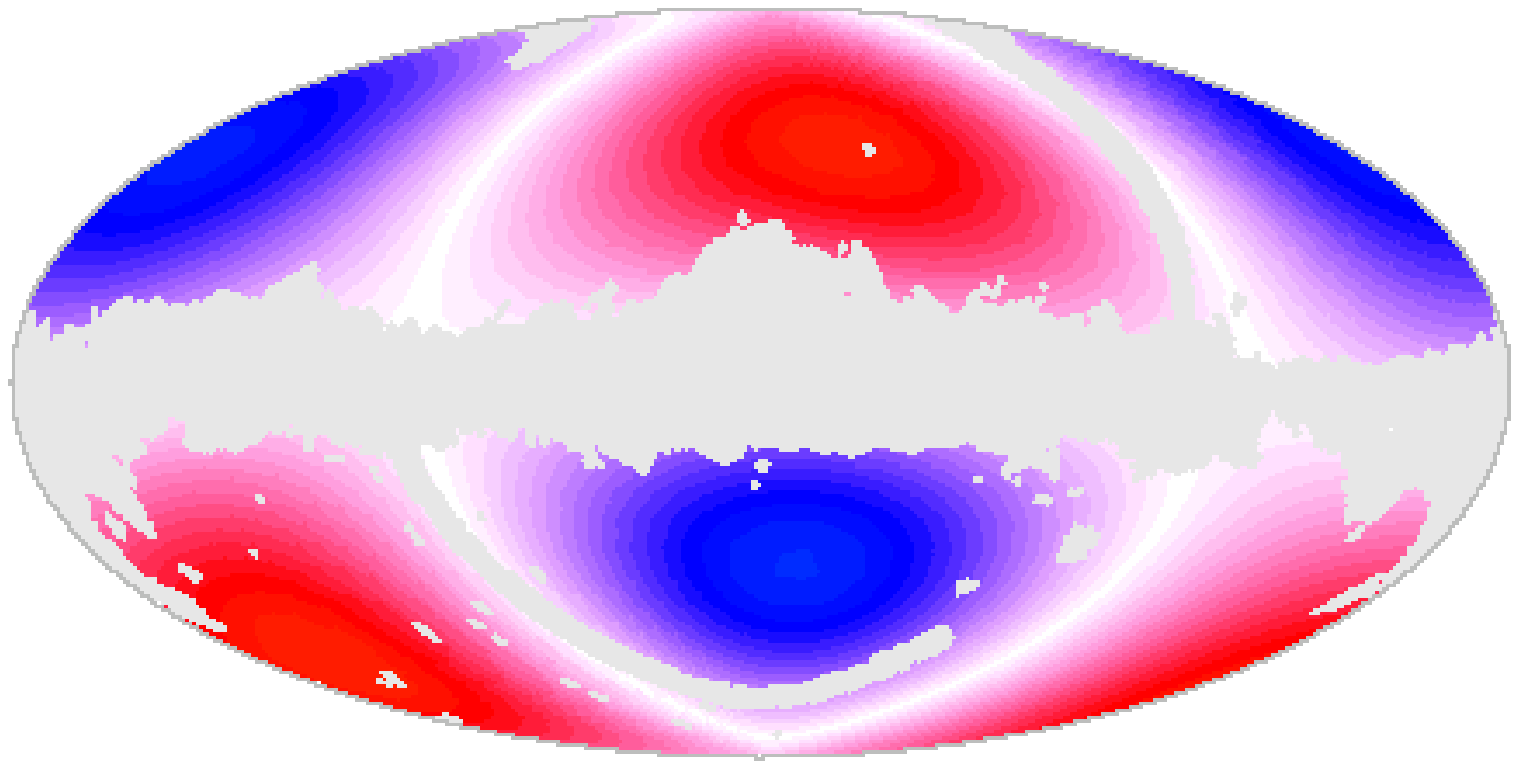}
    \includegraphics[scale=.36]{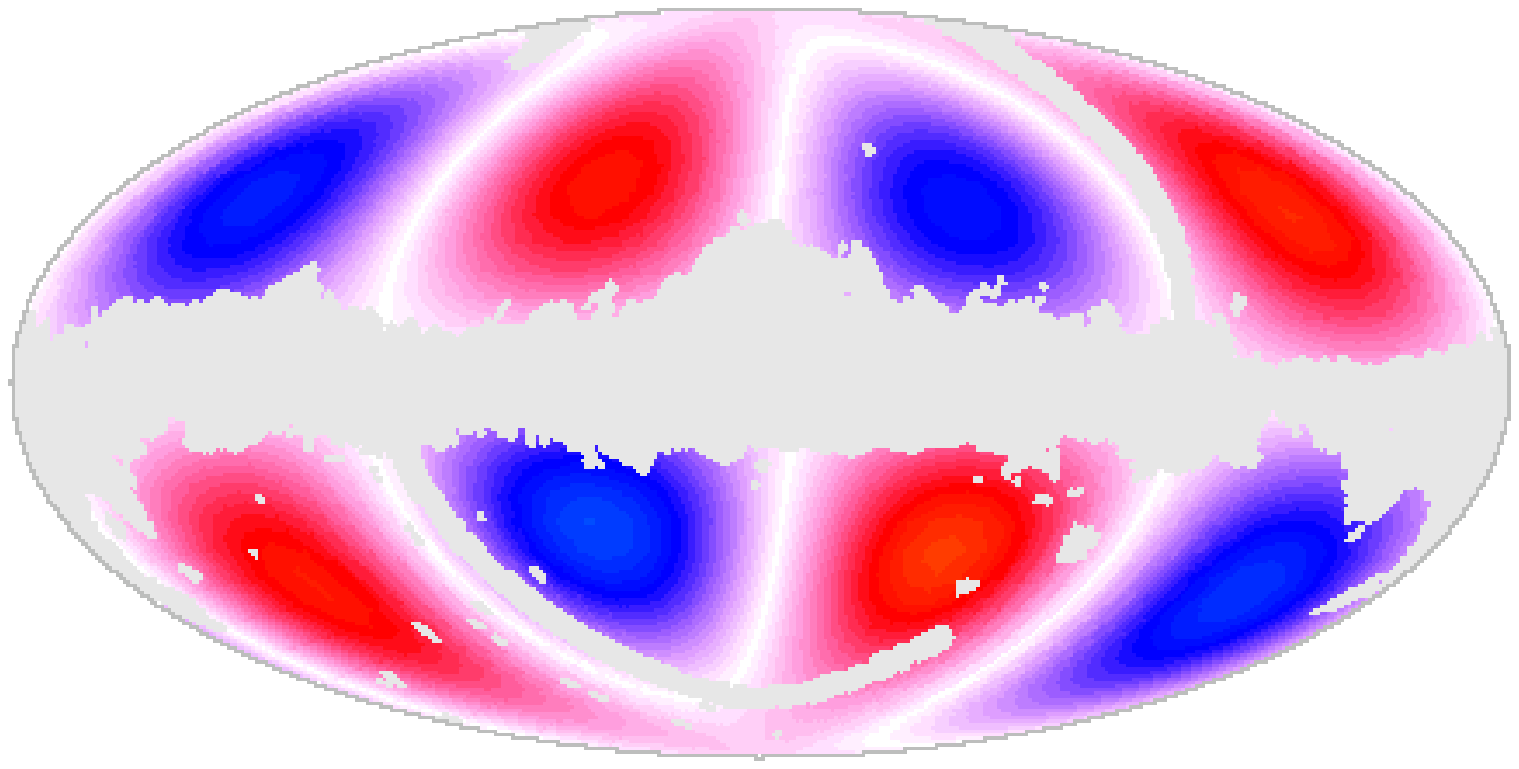}
    \includegraphics[scale=.36]{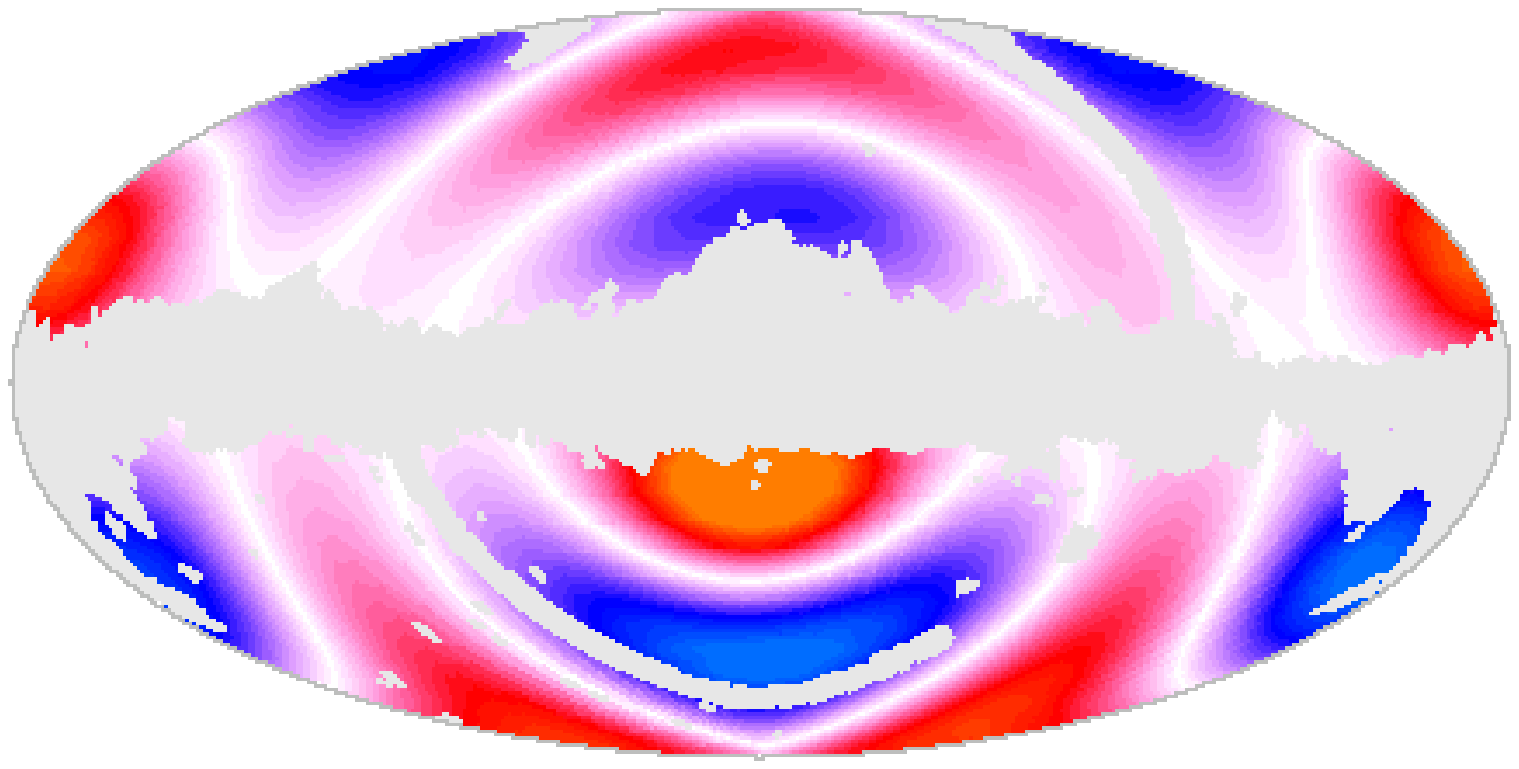}
    \includegraphics[scale=.36]{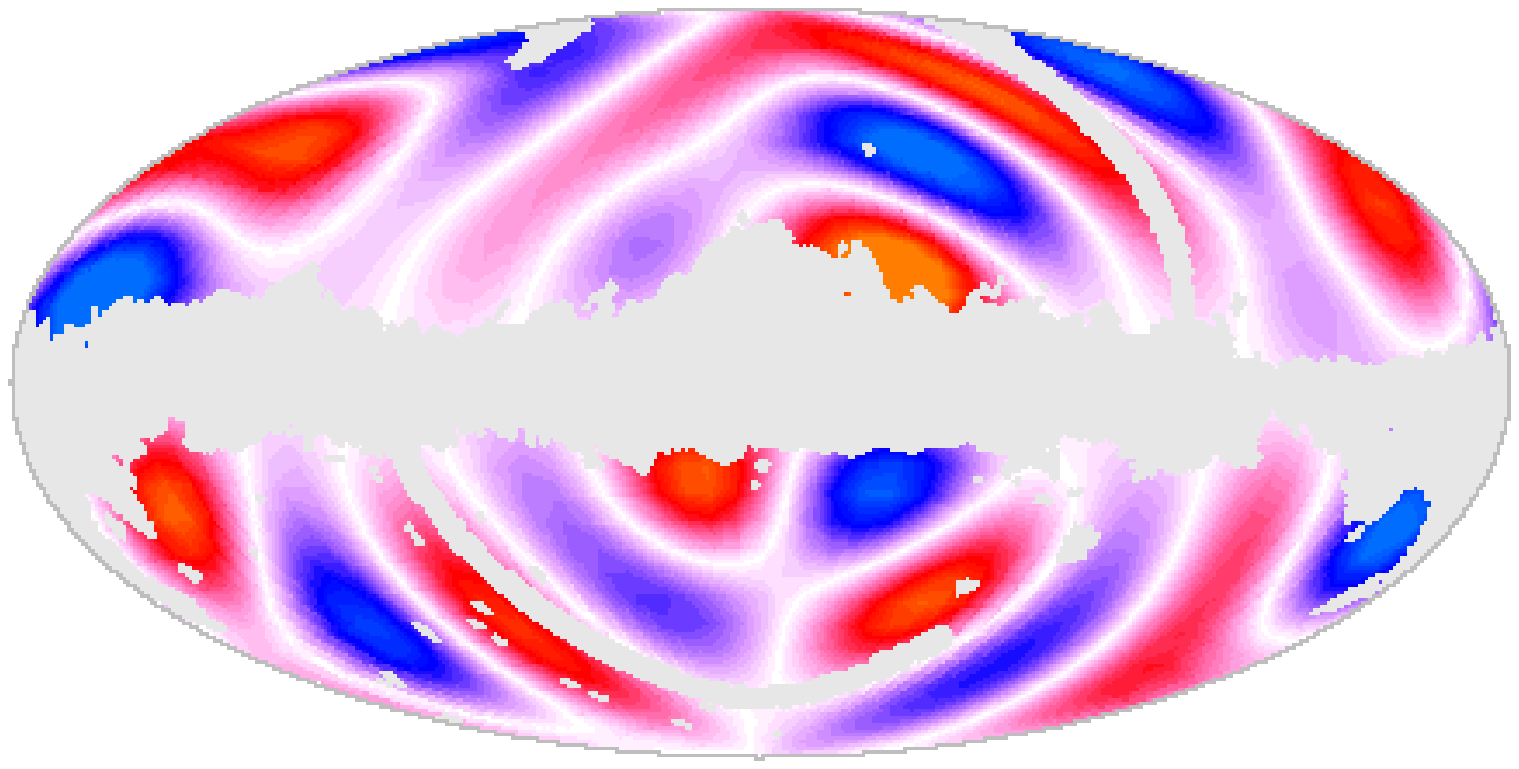}
    \includegraphics[scale=.36]{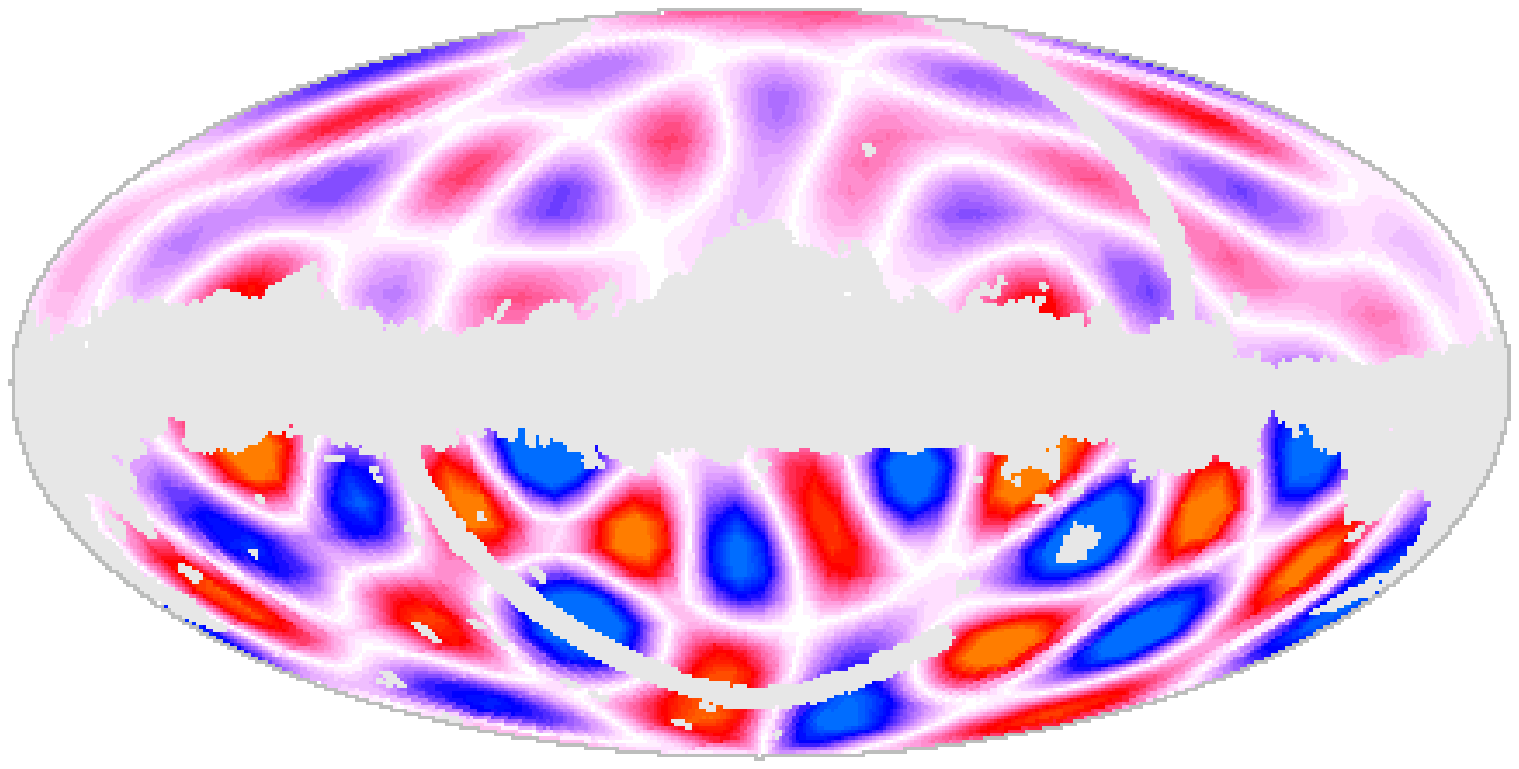}
    \includegraphics[scale=.36]{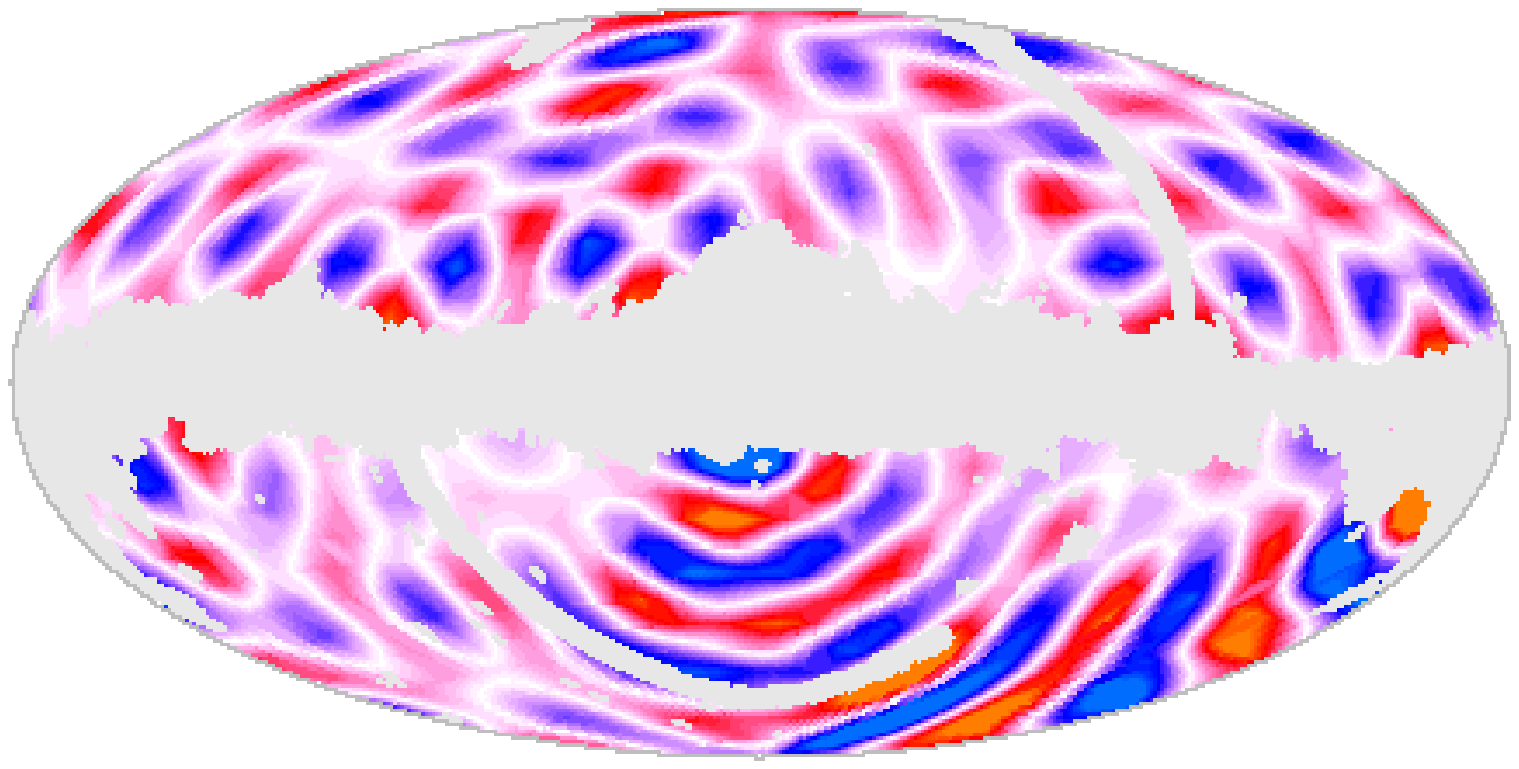}
    \end{center}
    \caption[1]{
    \label{mappscz}
Selection of angular Karhunen-Lo\`eve modes in the PSCz survey
with the high-latitude mask.
The projection is Hammer-Aitoff in galactic coordinates,
with the Galactic Centre at centre.
The top 4 modes are special,
while the rest are constructed from the KL procedure.
The angular modes are, from top left to bottom right:
mode 1, the (cut) monopole mode;
modes 2--4, the three (cut) dipole modes
(with small admixtures of cut monopole to make them orthogonal to the monopole);
then modes 5, 10, 20, 40, 80, and 160.
All modes are mutually orthogonal
over the unmasked part of the sky.
The modes are finite sums of harmonics up to $l = 39$.
Mode 5 (left middle)
is the (non-special)
KL mode containing the most information about large scale angular power.
The mode evidently ``knows about''
the PSCz angular mask:
the mode has low amplitude in masked regions of the survey,
and high amplitude in unmasked regions.
    }
    \end{figure}
}

\newcommand{\radialpsczhibfig}{
    \begin{figure}[t]
    \begin{center}
    \includegraphics[scale=.6]{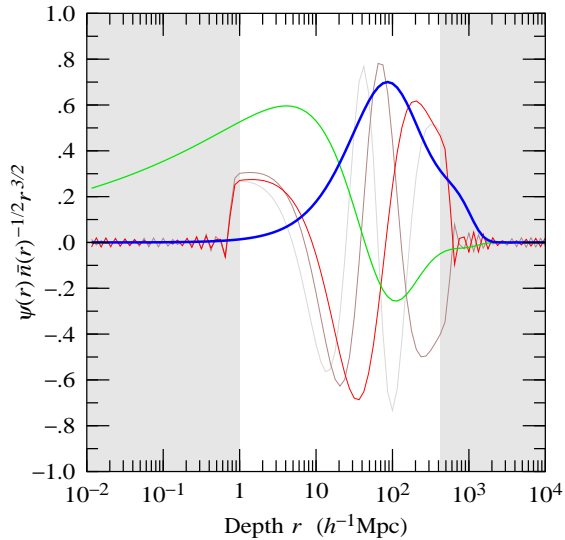}
    \end{center}
    \caption[1]{
    \label{radialpsczhib}
The first 5 radial Karhunen-Lo\`eve modes
associated with the (cut) monopole angular mode in the PSCz survey.
The first two radial modes are special:
mode 1 (thick blue) is the mean mode,
and mode 2 (medium green) is the Local Group (LG) motion mode
(with an admixture of the mean mode to make it orthogonal to the mean).
The remaining radial modes are constructed from the KL procedure.
The modes are all mutually orthogonal
over the (unshaded) interval from
$1 \, h^{-1} {\rm Mpc}$
to
$420 \, h^{-1} {\rm Mpc}$.
The modes are finite sums of logarithmic radial waves
(\S\protect\ref{redshift})
defined, to avoid aliasing, over the extended logarthmic interval
$10^{-2} \, h^{-1} {\rm Mpc}$
to
$10^4 \, h^{-1} {\rm Mpc}$.

Radial KL modes associated with other angular KL modes are similar
but not identical.
For non-special angular modes
(i.e.\ angular modes other than the cut monopole and dipole)
it is optional whether or not to force the first mode(s)
to be special.
Nowadays we tend to keep the mean but not the LG radial mode
in non-special angular modes.
Keeping the mean radial mode make it possible to test for
possible purely angular systematics,
such as might be associated with extinction.
    }
    \end{figure}
}

\newcommand{\aklfig}{
    \begin{figure}[t]
    \begin{center}
    \includegraphics[scale=.6]{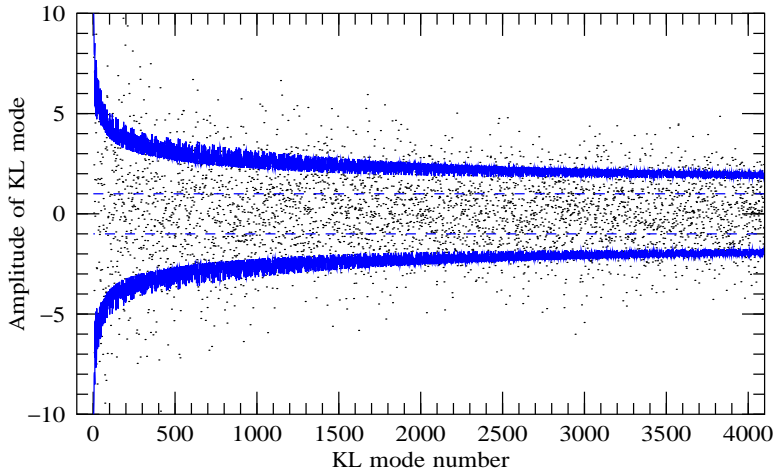}
    \end{center}
    \caption[1]{
    \label{akl}
Amplitudes of 4095 of the 4096 pseudo-Karhunen-Lo\`eve modes in the PSCz survey
(the missing mode is the mean mode, whose amplitude on this scale is huge).
The dots are the measured amplitudes,
which according to the prior are expected to be Gaussianly distributed
about zero, with expected standard deviation as given by the solid line.
The dashed line is the expected standard deviation from shot noise alone.
    }
    \end{figure}
}

\newcommand{\xilcontsfig}{
    \begin{figure}[t]
    \begin{center}
    \includegraphics[scale=.8,angle=270]{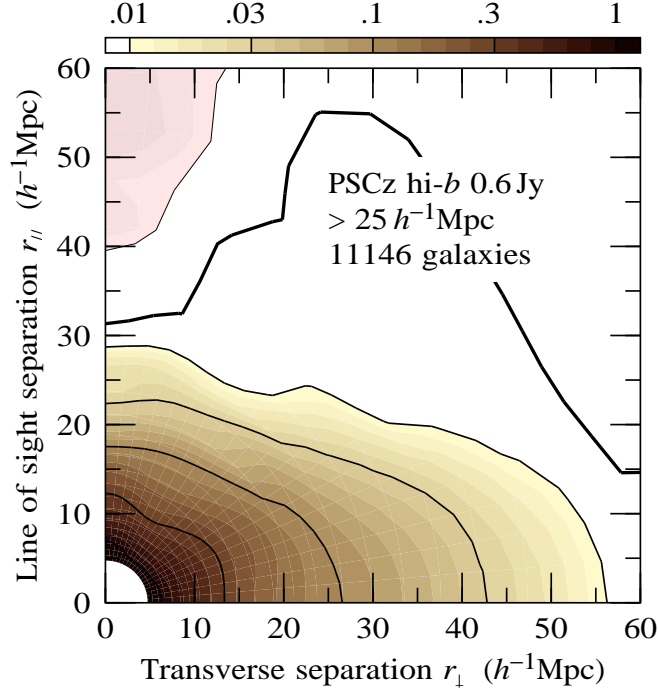}
    \end{center}
    \caption[1]{
    \label{xilconts}
Contour plot of the redshift space two-point correlation function
in the PSCz survey with the high galactic latitude angular mask.
Unlike the analysis discussed in the rest of this paper,
this plot assumes that redshift distortions are plane-parallel
(to which end only galaxies beyond $25 \, h^{-1} {\rm Mpc}$ are included).
The expected linear squashing effect is plainly visible,
while nonlinear fingers-of-god show up as a mild extension
along the line-of-sight axis.
Thin, medium, and thick contours represent
negative, positive, and zero values respectively.
The correlation function has been smoothed over pair separation
$r = ( r_\perp^2 + r_\parallel^2 )^{1/2}$
with a tophat window of width $0.2$~dex,
and over angles $\theta = \tan^{-1} ( r_\perp / r_\parallel )$
to the line of sight with a Gaussian window
with a 1$\sigma$ width of $10^\circ$.
From
Hamilton, Tegmark \& Padmanabhan (2000) \cite{HTP00}.
    }
    \end{figure}
}

\newcommand{\picfig}{
    \begin{figure}[t]
    \begin{center}
    \includegraphics[scale=.8]{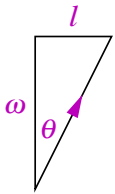}
    \end{center}
    \caption[1]{
    \label{pic}
The dimensionless log frequency $\omega$
is the radial analogue of the dimensionless spherical harmonic number $l$.
A logarithmic spherical wave $Z_{\omega lm}$
with log frequency $\omega$ and harmonic number $l$
has wavevector angled effectively at $\theta = \tan^{-1} (l/\omega)$
to the line-of-sight.
    }
    \end{figure}
}

\newcommand{\sqrtdemofig}{
    \begin{figure}[t]
    \begin{center}
    \includegraphics[scale=.6]{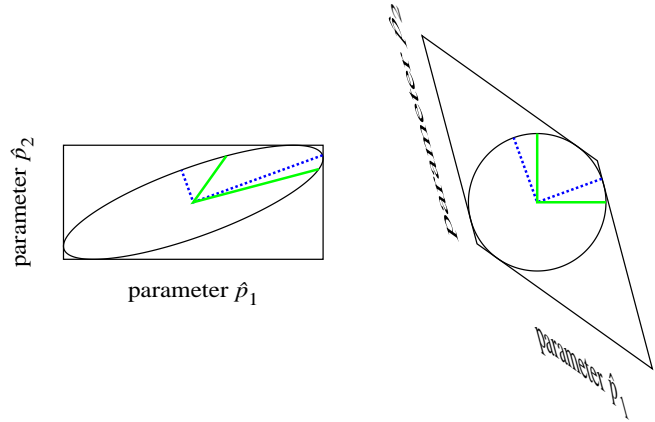}
    \end{center}
    \caption[1]{
    \label{sqrtdemo}
A vector $\hat\pvec = (\hat p_1 , \hat p_2)$
of correlated parameter estimators.
The thick
(blue)
dashed lines represent the Principal Component Decomposition of $\hat p_\alpha$,
the eigenvectors of their covariance matrix.
They are uncorrelated.
The thick
(green)
solid lines represent $F^{1/2} \hat\pvec$,
the parameter estimates decorrelated with the square root of the Fisher matrix.
They are also uncorrelated.
The diagram on the right is the same as that on the left,
but stretched out along the minor axis of the error ellipse,
so that the error ellipse becomes an error circle.
Any two vectors that are orthogonal on the error circle are uncorrelated.
    }
    \end{figure}
}

\newcommand{\sqrtffig}{
    \begin{figure}[t]
    \begin{center}
    \includegraphics[scale=.67]{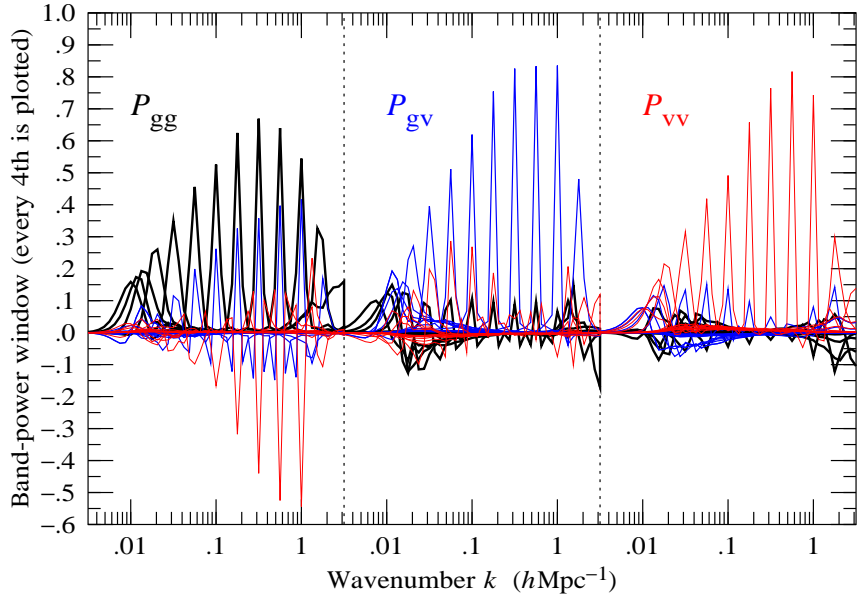}
    \end{center}
    \caption[1]{
    \label{sqrtf}
Band-power windows
for the disentangled, decorrelated power spectra
measured from the PSCz survey
for (left) the galaxy-galaxy power spectrum $P_{\rm gg}$,
(middle) the galaxy-velocity power spectrum $P_{\rm gv}$,
and (right) the velocity-velocity power spectrum $P_{\rm vv}$.
Each band-power includes contributions from all three power spectra,
the (thick black) ${\rm gg}$,
(medium blue) ${\rm gv}$,
and (thin red) ${\rm gg}$ powers,
but the off-type contributions cancel, according to the prior.
For example, in the left panel, the contributions to the
${\rm gg}$ band-power
from each of the ${\rm gv}$ and ${\rm vv}$ powers
should sum to zero, if the prior is correct.
    }
    \end{figure}
}

\newcommand{\xikfullfig}{
    \begin{figure}[t]
    \begin{center}
    \includegraphics[scale=.65]{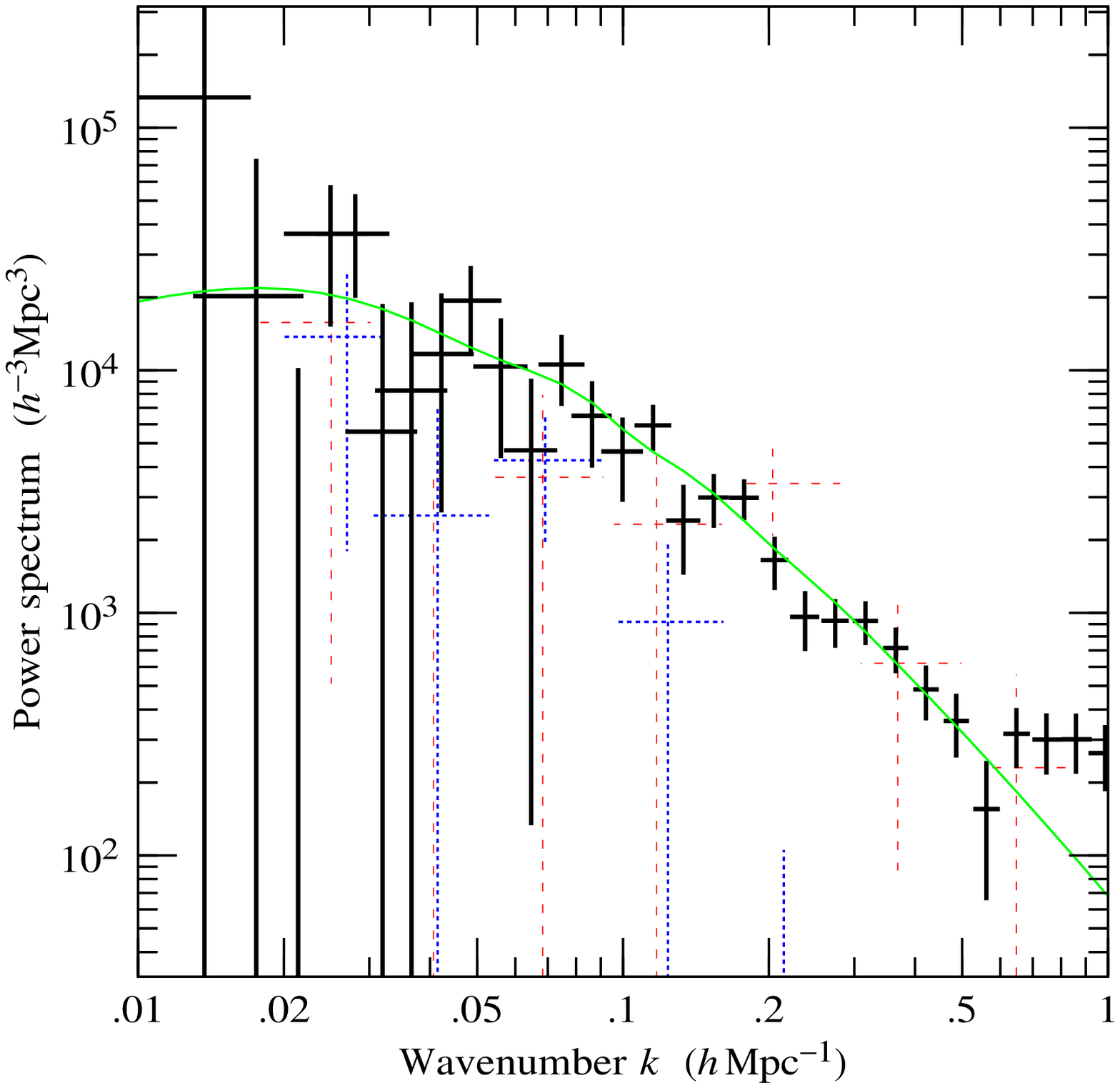}
    \end{center}
    \caption[1]{
    \label{xikfull}
Points with error bars show the
decorrelated power spectra measured from the PSCz survey:
(thick black) the galaxy-galaxy power spectrum $P_{\rm gg}$,
(medium blue dotted) the galaxy-velocity power spectrum $P_{\rm gv}$,
and
(thin red dashed) the velocity-velocity power spectrum $P_{\rm gv}$.
The ${\rm gv}$ and ${\rm vv}$ powers are plotted with coarser binning
because they are noisier than the ${\rm gg}$ power.
Each point is plotted at the median wavenumber of its band-power window;
the horizontal error bar gives the FWHM of the band-power window.
The smooth solid (green) line running through the galaxy-galaxy power spectrum
is the linear concordance $\varLambda$CDM model of
Tegmark, Zaldarriaga \& Hamilton (2001)
\protect\cite{TZH01}.
    }
    \end{figure}
}

\hyphenpenalty=3000

\maketitle

\section*{Abstract}
This second paper of two companion papers
on the estimation of power spectra specializes to the topic
of estimating galaxy power spectra at large, linear scales
using maximum likelihood methods.
As in the first paper,
the aims are pedagogical,
and the emphasis is on concepts rather than technical detail.
The paper covers most of the salient issues, including
selection functions,
likelihood functions,
Karhunen-Lo\`eve compression,
pair-integral bias,
Local Group flow,
angular or radial systematics
(arising for example from extinction),
redshift distortions,
quadratic compression,
decorrelation,
and disentanglement.
The procedures are illustrated with results from the
{\it IRAS\/} PSCz survey.
Most of the PSCz graphics included in this paper
have not been published elsewhere.

\section{Introduction}

This paper addresses the problem
of estimating galaxy power spectra at linear scales
using maximum likelihood methods.
As discussed in Paper~1,
the power spectrum is the most important statistic that
can be measured from large scale structure (LSS),
and, again as elaborated in Paper~1,
maximum likelihood has a special status in statistics:
if a best method exists, then it is the maximum likelihood method
(Tegmark, Taylor \& Heavens 1997 \cite{TTH97}).

Fisher, Scharf \& Lahav (1994) \cite{KSL94} were the first to apply
a likelihood approach to large scale structure.
Heavens \& Taylor (1995) \cite{HT95}
may be credited with accomplishing the first likelihood analysis
designed to retain as much information as possible at linear scales.
The primary goal of
both Fisher et al \cite{KSL94}
and Heavens \& Taylor \cite{HT95}
was to measure the linear redshift distortion parameter
$\beta \approx \varOmega_m^{4/7} / b$
from the {\it IRAS\/} 1.2~Jy survey.
Shortly thereafter,
Ballinger, Heavens \& Taylor (1995) \cite{BHT95}
extended the analysis to include a measurement of the real space
(as opposed to redshift space) linear power spectrum
of the 1.2~Jy survey
in several bins of wavenumber.

Several authors have built on and applied the
Heavens \& Taylor's (1995) \cite{HT95} method.
Tadros et al (1999) \cite{TBTHESFKMMORSW99}
and
Taylor et al (2001) \cite{TBHT01}
applied improved versions of the method to measure
the power spectrum and redshift distortions in
the {\it IRAS\/} Point Source Catalogue redshift survey
(PSCz; Saunders et al 2000 \cite{S00}).
More recently,
Percival et al (2004) \cite{P04}
applied the method to the final version of the
2 degree Field Galaxy Redshift Survey (2dFGRS; Colless et al 2003 \cite{C03}).

The present paper aims at a pedagogical
presentation of the various issues involved
in carrying out a maximum likelihood analysis of
the galaxy power spectrum and its redshift distortions.
The paper is based largely on our own work over the last several years
(Hamilton, Tegmark \& Padmanabhan 2000 \cite{HTP00};
Padmanabhan, Tegmark \& Hamilton 2001 \cite{PTH01};
Tegmark, Hamilton \& Xu 2002 \cite{THX02};
Tegmark et al 2004 \cite{T04}).
The procedures are illustrated here mainly with measurements
from the {\it IRAS\/} PSCz survey (Saunders et al 2000 \cite{S00}).
The measurements are based on the work reported by
Hamilton, Tegmark \& Padmanabhan (2000) \cite{HTP00},
but most of the graphics in the present paper
appear here for the first time.

The paper starts, \S\ref{results},
by showing the final real space linear power spectra
measured by the methods described herein
from the PSCz, 2dF, and SDSS surveys.
The remainder of the paper is organized into sections
each dealing with a specific aspect of measuring the linear power spectrum:
\S\ref{selectionfunction}
selection function;
\S\ref{nonlinear}
linear vs.\ nonlinear regimes;
\S\ref{gaussianlik}
Gaussian likelihood function;
\S\ref{numerical}
numerical obstacle;
\S\ref{KL}
Karhunen-Lo\`eve compression;
\S\ref{pairintegral}
removing pair-integral bias;
\S\ref{LGflow}
Local Group flow;
\S\ref{systematics}
isolating angular and radial systematics;
\S\ref{redshift}
redshift distortions and logarithmic spherical waves;
\S\ref{quadratic}
quadratic compression;
\S\ref{decorrelation}
decorrelation;
and
\S\ref{disentanglement}
disentanglement.
Finally, 
\S\ref{conclusion}
summarizes the onclusions.

\xikfig

\section{Results}
\label{results}

Figure~\ref{xik},
taken from Tegmark et al (2004) \cite{T04},
shows the linear galaxy power spectra
measured, by the methods described in this paper,
from the PSCz
(Saunders et al 2000 \cite{S00}),
2dF (Colless et al 2003 \cite{C03}),
and SDSS (York et al 2000 \cite{Y00})
surveys.

Two important points to note about these power spectra are,
first,
that the power spectra have redshift distortions removed
(\S\ref{redshift}),
and are therefore real space power spectra,
and second,
that the power spectra have been decorrelated
(\S\ref{decorrelation}),
so that each point with its error bar represents
a statistically independent object.
Each point represents not the power at a single wavenumber,
but rather the power in a certain well-defined band
(\S\ref{disentanglement}).


\section{Selection Function}
\label{selectionfunction}

A prerequisite for measuring galaxy power spectra,
whether at linear or nonlinear scales,
is to measure the angular and radial selection functions of a survey.
All I really want to say here is that selection functions
do not grow on trees,
but require a lot of sometimes unappreciated hard work to measure.

As regards the angular selection function,
one thing that can help cut down the work and improve accuracy
is a software package \mangle\ that we recently published
(Hamilton \& Tegmark 2004 \cite{HT04}).
Max tells me that \mangle\ has become the de facto software
with which the SDSS team are characterizing the angular selection function
of the SDSS.

Measuring the radial selection function is similarly onerous.
The principal goal of all the methods is to separate
out the smooth radial variation of the selection function
from the large variations in galaxy density caused by galaxy clustering.
Invariably, the essential assumption made to effect this separation
is that the galaxy luminosity function is a universal function,
independent of position.
One of the best papers
(in terms of content, as opposed to comprehensibility)
on the subject remains the seminal paper by
Lynden-Bell (1971) \cite{LB71},
who applied a maximum likelihood method.
Other papers include:
Sandage, Tammann \& Yahil (1979) \cite{STY79},
Cho{\l}oniewski (1986, 1987) \cite{Chol86,Chol87},
Binggeli, Sandage \& Tammann (1988) \cite{BST88},
Efstathiou, Ellis \& Peterson (1988) \cite{EEP88},
SubbaRao et al (1996) \cite{SCSK96},
Heyl et al (1997) \cite{HCEB97},
Willmer (1997) \cite{W97},
Tresse (1999) \cite{T99}.

\section{Linear vs.\ Nonlinear Regimes}
\label{nonlinear}

There is a big difference between the linear and nonlinear regimes
of galaxy clustering,
and it makes sense to measure the power spectrum
using different techniques in the two regimes.

At linear scales,
it is legitimate to assume a much tighter prior than
at nonlinear scales.
On the other hand,
at linear scales
there are far fewer modes
than at nonlinear scales.

At linear scales,
you can reasonably assume (with various degrees of confidence):
\begin{itemize}
\item[$\bullet$]
Gaussian fluctuations;
\item[$\bullet$]
redshift distortions conform to the linear Kaiser (1987) \cite{K87} model;
\item[$\bullet$]
linear bias between galaxies and matter.
\end{itemize}

At nonlinear scales,
all of the above assumptions are false,
and it would be an error to assume that they are true.
There is however one useful assumption
that is a better approximation at nonlinear scales than at linear scales:
\begin{itemize}
\item[$\bullet$]
redshift distortions are plane-parallel (the distant observer approximation).
\end{itemize}

The remainder of this paper
devotes itself to the problem of measuring power at linear scales.

\section{Gaussian Likelihood Function}
\label{gaussianlik}

Measuring the power spectrum of a galaxy survey at linear scales
starts with the fundamental prior assumption that
the density field is Gaussian.
With this assumption, one has the luxury of being able to write
down an explicit Gaussian likelihood function
\begin{equation}
\label{lik}
  {\cal L} \propto
  {1 \over \left| C \right|^{1/2}}
  \exp \left( - \frac{1}{2} C_{ij}^{-1} \delta_i \delta_j \right)
\end{equation}
where $\delta_i$ is a vector of measured overdensities,
$C_{ij}$ is their covariance matrix (part of the prior)
and $\left| C \right|$ is the determinant of the covariance matrix.
Normally,
the covariance matrix is assumed to be a sum of two parts,
a cosmic part and a shot noise part -- see \S\ref{KL} below.

It should be emphasized that the assumption of Gaussian fluctuations
is {\em not\/} valid at nonlinear scales,
and it would be wrong to assume that the likelihood~(\ref{lik})
holds at nonlinear scales.
If you use the likelihood~(\ref{lik}) at nonlinear scales,
then you will substantially underestimate the true error bars
on the power spectrum.

\section{Numerical Obstacle}
\label{numerical}

Perhaps the biggest single obstacle to overcome
in carrying out a Gaussian likelhood analysis is a numerical one,
the limit on the size of the covariance matrix $C$
that the numerics can deal with.
Suppose that you have $N$ modes in the likelihood function.
Manipulating the $N \times N$ covariance matrix $C$
is typically an $N^3$ process.
Thus doubling the number of modes takes $2^3 = 8$ times as much computer time.

To make matters worse,
consider the fact that the number of modes in a survey
increases as roughly the cube of the maximum wavenumber
(smallest scale) to which you choose to probe, $N \sim k_{\max}^3$.
The numerical problem thus scales as $k_{\max}^9$.
In other words, if you want to push to half the scale,
by doubling $k_{\max}$,
then you need $2^3 = 8$ times as many modes,
and $2^9 = 512$ times as much computer time.

Fortunately,
the linear regime is covered with a tractable number of modes.
The boundary between linear and nonlinear regimes is at
$k \approx 0.3 \, h \, {\rm Mpc}^{-1}$.
If you prefer to remain safely in the linear regime,
you may prefer to stick to
$k \lesssim 0.1 \, h \, {\rm Mpc}^{-1}$,
as did Heavens \& Taylor (1995) \cite{HT95}.
Max Tegmark and I typically have used about 4000 modes
(a week of computer time on a workstation),
which took us to
$k \approx 0.25 \, h \, {\rm Mpc}^{-1}$
in the PSCz survey,
and
$k \approx 0.15 \, h \, {\rm Mpc}^{-1}$
in the 2dF 100k
and SDSS
surveys.

The previous paragraph would seem to suggest that 4000 modes is fine,
but I have to confess that although our coverage of information is close
to 100\% at the largest scales, we lose progressively more information
at smaller scales.
To catch all the information available at say
$k \approx 0.1 \, h \, {\rm Mpc}^{-1}$,
we should really push to $10^4$ or $10^5$ modes.

How does one deal with the numerical limitation of being able to use only
a finite number of modes?
One thing is to make sure that your code is as fast as can be.
Several years ago
Max Tegmark taught me some tricks
that can speed up matrix manipulations by 1 or 2 orders of magnitude
(without which our code would take a year to run instead of a week).
It helps to time the various steps in the code,
and to oil the bottlenecks.

The other big thing to get around the numerical limitations
is to use the Karhunen-Lo\`eve technique
to compress the information of interest
-- power at linear scales -- into a modest number of modes.

\mappsczfig

\radialpsczhibfig

\aklfig

\section{Karhunen-Lo\`eve Compression}
\label{KL}

The idea of Karhunen-Lo\`eve (KL) compression
is to keep only the highest signal-to-noise modes
in the likelihood function.
This idea was first proposed by
Vogeley \& Szalay (1996) \cite{VS96}
for application to LSS.

Suppose that the covariance matrix $C$
is a sum of a signal $S$ (the cosmic variance)
and noise $N$ (the shot noise)
\begin{equation}
  C = S + N
  \;.
\end{equation}
Prewhiten the covariance matrix,
that is, transform the covariance matrix so that the noise matrix
is the unit matrix:
\begin{equation}
  N^{-1/2} C N^{-1/2} = N^{-1/2} S N^{-1/2} + 1
\end{equation}
where the $1$ on the RHS is to be interpreted as the unit matrix.
Diagonalize the prewhitened signal:
\begin{equation}
  N^{-1/2} S N^{-1/2} = O \varLambda O^\transpose
\end{equation}
where $O$ is an orthogonal matrix and $\varLambda$ is diagonal.
Since the unit matrix remains the unit matrix under any diagonalization,
the covariance matrix is
\begin{equation}
  N^{-1/2} C N^{-1/2} = O ( \varLambda + 1 ) O^\transpose
  \;.
\end{equation}
The resulting eigenmodes,
the columns of the orthogonal matrix $O$,
are \textbf{Karhunen-Lo\`eve},
or \textbf{signal-to-noise}, eigenmodes,
with eigenvalues $\lambda$ equal to the signal-to-noise ratio of each mode.

The KL procedure thus parcels the information in a signal into
a discrete set of modes ordered by their signal-to-noise.
This is definitely a very neat trick.

However, there is a big drawback to KL compression,
which is that you typically want to extract $N$ modes from
a much larger pool of ${\cal N} \gg N$ modes
-- that's why it's called compression.
And that requires diagonalizing a large ${\cal N} \times {\cal N}$ matrix.
But the whole point of KL compression
is to avoid having to mess with a large matrix.

Fortunately, there is a way out of this loop.
The first point to note is that
the thing of interest is the ensemble of KL modes,
not each mode individually (though they are cute to look at
-- see Figures~\ref{mappscz} \& \ref{radialpsczhib}).
Thus the KL modes do not have to be perfect.
The second point is that what you call ``signal'' is up to you.
In the case under consideration the signal is
``the power spectrum at linear scales'',
which is actually not a single signal but a whole suite of signals.

Our strategy to solve the KL crunch
is to compress first into a set of angular KL modes
(Figure~\ref{mappscz});
and then within each angular KL mode to compress
into a set of radial KL modes
(Figure~\ref{radialpsczhib}).
The result is a set of ``pseudo-Karhunen-Lo\`eve'' (PKL) modes
each of which is the product of an angular and a radial profile.
The PKL modes are not perfect,
but they cover the relevant subspace of Hilbert space without gaps,
and that's all we need.

Since the goal is to measure power at linear scales,
we choose the ``signal'' to be not a realistic power spectrum,
but rather an artificial power spectrum $P(k) \propto k^{-2.5}$
which increases steeply to large scales.
Thus our procedure favours modes that are sensitive to power at large scales;
but a low-noise small scale mode
can beat out a noisy large scale mode.

Typically,
we start with a few thousand angular modes in spherical harmonic space,
and apply KL diagonalization to these angular modes.
We then march through each angular KL mode one by one.
Within each angular KL mode,
we resolve the radial direction into several hundred
logarithmic spherical waves,
and apply KL diagonalization to those.
We keep a running pool of the best 4000 modes so far.
There is no need to go through all the angular KL modes.
The later angular KL modes contain little information,
and when we have gone through 10 successive angular KL modes
and found no new mode good enough to make it into the pool of best modes,
we stop.
The procedure effectively compresses $10^5$--$10^6$ modes into 4000,
but remains well within the capabilities of a modern workstation.

Figure~\ref{akl}
shows the amplitudes $x_i$ of 4095 of 4096 PKL modes
(the omitted mode is the mean, equation~(\ref{psimean}),
whose amplitude, equation~(\ref{xmean}),
is huge compared to all other modes)
measured in the PSCz survey.
According to the prior,
the amplitudes should be Gaussianly distributed about zero
(excepting the mean mode),
with variances given by the diagonal elements
$C_{ii} = \langle \varDelta x_i^2 \rangle$
of the (prior) covariance matrix of PKL modes.
Indeed, the measured amplitudes are consistent with this prior.

\section{Removing Pair-Integral Bias}
\label{pairintegral}

You are undoubtedly familiar with the notion that if you measure
both the mean and the variance from a data set,
then the measurement of the variance will be biassed low.
The usual fix up is, if you have $N$ independent data,
to divide the sum of the squared deviations
by $N - 1$ rather than $N$.

Applied to LSS,
this bias is known in the literature as the ``pair integral constraint''
(the observed number of neighbours of a galaxy in a survey
equals the number of galaxies in the survey minus one).

The simplistic procedure of dividing by $N-1$ instead of $N$
does not work in LSS,
but Fisher et al (1993) \cite{FDSYH93}
pointed out a delightfully simple trick
that does completely solve the problem.
The Fisher et al trick is to isolate the mean,
the selection function $\nbar(\rvec)$,
into a single ``mean mode'',
and to make all other modes orthogonal to the mean.

In the present context,
let $\psi_i(\rvec)$ denote a PKL mode.
The observed amplitude $x_i$ of a PKL mode is defined to be
\begin{equation}
  x_i
  \equiv
  \int \psi_i(\rvec) {n(\rvec) \over \nbar(\rvec)} \, \ddd r
  =
  \sum_{{\rm galaxies}~g} {\psi_i(\rvec_g) \over \nbar(\rvec_g)}
  \;.
\end{equation}
The mean mode
$\psi_1(\rvec)$
is defined to be the mode whose shape is the mean
\begin{equation}
\label{psimean}
  \psi_1(\rvec) \equiv \nbar(\rvec)
  \;.
\end{equation}
The amplitude of the mean mode is
\begin{equation}
\label{xmean}
  x_1 = \int n(\rvec) \, \ddd r
  = N_{\rm gal}
\end{equation}
the number of galaxies in the survey.
Fisher et al's (1993) \cite{FDSYH93}
trick is to arrange all modes other than the mean
to be orthogonal to the mean
\begin{equation}
  \langle x_i \rangle = \int \psi_i(\rvec) \, \ddd r = 0
  \;.
\end{equation}

This trick ensures that the amplitudes of all modes except the mean mode
are unaffected (to linear order, anyway) by uncertainty in the mean.
The resulting power spectrum is unbiassed by the pair integral constraint.
Neat!

The observed amplitude of the mean mode is used
in computing the maximum likelihood normalization of the selection function,
but is then discarded from the analysis,
because it is impossible to measure the fluctuation of the mean mode,
just as it is impossible to measure the fluctuation of the monopole mode
in the CMB.

\section{Local Group Flow}
\label{LGflow}

The motion of the Local Group (LG) induces a dipole in the density distribution
around it
(Hamilton 1998 \cite{H98}, eq.~(4.42))
(or rather,
since the LG is going with the flow,
the motion of the LG removes the dipole present in the CMB frame).
Although the LG mode is a single mode,
we choose to project the effect, along with the mean mode,
into a set of 8 modes,
whose angular parts are (cut) monopole and dipole ($1 + 3 = 4$ modes),
illustrated in the top four panels of Figure~\ref{mappscz},
and whose radial parts are the mean mode $\nbar(r)$
and the LG radial mode
$\partial \left[ r^2 \nbar(r) \right] / r^2 \partial r$
($1 + 1 = 2$ modes),
illustrated in Figure~\ref{radialpsczhib}.

Since the motion of the LG through the CMB is known
(Bennett et al 2003 \cite{B03};
Courteau \& van den Bergh 1999 \cite{CvdB99};
Lineweaver et al 1996 \cite{LTSKBL96}),
the amplitudes of the LG modes
can be corrected for this motion,
and included in the analysis.
This is unlike the CMB,
where the fluctuations of the dipole modes cannot be measured
separately from the motion of the Sun,
and must be discarded.

\section{Isolating Angular and Radial Systematics}
\label{systematics}

A similar trick can be used to isolate other potential problems
into specific modes or sets of modes.

For example,
possible angular systematics,
associated for example
with uncertainties in angular extinction across the sky,
can be projected into a set of purely angular modes
(whose radial part is the mean radial mode).
Other modes should be unaffected (to linear order)
by such systematics,
because they are orthogonal to purely angular variations.
If a systematic effect arising from extinction were present,
then it would show up as a systematic enhancement of power
in the purely angular modes.

Similarly,
possible radial systematics,
associated perhaps with uncertainties in the radial selection function,
or with evolution as a function of redshift,
can be projected into a set of purely radial modes
(whose angular part is the cut monopole).

We did not project out purely angular or radial modes
in our PSCz analysis,
but we have been doing this in our more recent work
(Tegmark, Hamilton \& Xu 2002 \cite{THX02};
Tegmark et al 2004 \cite{T04}).

\xilcontsfig

\section{Redshift Distortions and Logarithmic Spherical Waves}
\label{redshift}

Large scale coherent motions towards overdense regions
induce a {\bf linear squashing} effect on the correlation function
of galaxies observed in redshift space,
as illustrated in
Figure~\ref{xilconts}
for the PSCz survey.
At smaller scales,
collapse and virialization
gives rise to so-called {\bf fingers-of-god},
visible in Figure~\ref{xilconts}
as a mild extension of the correlation function
along the line-of-sight axis.



Kaiser (1987) \cite{K87} first pointed out the celebrated result that,
at linear scales, and in the plane-parallel (distant observer) approximation,
the Fourier amplitude $\delta^{(s)}(\kvec)$
of galaxies in redshift space is amplified
over the Fourier amplitude $\delta(\kvec)$ of mass in real space
by a factor $b + f \mu_\kvec^2$
\begin{equation}
\label{dkaiser}
  \delta^{(s)}(\kvec) = ( b + f \mu_\kvec^2 ) \delta(\kvec)
\end{equation}
where $\mu_\kvec = \zhat . \khat$ is the cosine of the angle between
the wavevector $\kvec$ and the line of sight $\zhat$,
the quantity $b$ is the linear galaxy-to-mass bias,
and
$f \approx \varOmega_m^{4/7}$
is the dimensionless linear growth rate of fluctuations.
It follows immediately from Kaiser's formula that,
again at linear scales, and in the plane-parallel approximation,
the redshift space galaxy power spectrum $P^s(\kvec)$ is amplified over the
real space matter power spectrum $P(\kvec)$ by
\begin{equation}
\label{Pkaiser}
  P^{(s)}(\kvec) = ( b + f \mu_\kvec^2 )^2 P(\kvec)
  \;.
\end{equation}
Translated from Fourier space into real space,
Kaiser's formula predicts the large scale squashing effect
visible in Figure~\ref{xilconts}.

For the linear likelihood analysis being considered here,
the assumption of linear redshift distortions is fine,
but the plane-parallel approximation is not adequate.
Kaiser (1987) \cite{K87}, already in his original paper,
presented formulae for radial redshift distortions.
A pedagogical derivation can be found in the review by Hamilton (1998) \cite{H98}.
Unfortunately,
when the radial character of the redshift distortions is taken into account,
the formula for the amplification of modes ceases to be anything like as simple
as Kaiser's formula~(\ref{dkaiser}).
Indeed, the full, correct formula is more complicated in Fourier space
than in real space.

\picfig

In our pipeline,
we use logarithmic spherical waves as the fundamental basis with
respect to which we express PKL modes,
in part because radial redshift distortions take a simple form
in that basis, as first pointed out by Hamilton \& Culhane (1996) \cite{HC96}.
Logarithmic spherical waves are products of logarithmic radial waves
$\e^{\im \omega \ln r}$ and spherical harmonics $Y_{lm}(\rhat)$
\begin{equation}
\label{logwaves}
  Z_{\omega lm}(\rvec)
  =
  \e^{\im \omega \ln r} Y_{lm}(\rhat)
\end{equation}
and are eigenmodes of the complete set of commuting Hermitian operators
\begin{equation}
  - \im \left( {\partial \over \partial \ln r} + {3 \over 2} \right)
  \;, \quad
  L^2
  \;, \quad
  L_z
  \;.
\end{equation}
If you wonder why no one ever told you about logarithmic spherical waves
in quantum mechanics, so do I!
They are beautiful things.
For example, the logarithmic radial frequency $\omega$
is the radial analogue of the angular harmonic number $l$
(see Figure~\ref{pic})
\begin{equation}
  \omega \leftrightarrow {\rm radial}
  \quad {\rm as} \quad
  l \leftrightarrow {\rm angular}
\end{equation}
something that everyone ought to know.

In logarithmic spherical wave space,
the overdensity
$\delta_{\omega lm}^{(s)}$
of galaxies in redshift space
is related to the overdensity
$\delta_{\omega lm}$
of mass in real space by
\begin{equation}
\label{dolms}
  \delta_{\omega lm}^{(s)}
  =
  \left[ b + f
    {(\im \omega + 1/2) (\im \omega - 1/2) - \alpha(r) (\im \omega - 1/2)
    \over
    (\im \omega + l - 1/2) (\im \omega - l - 3/2)}
  \right] \delta_{\omega lm}
  \;.
\end{equation}
Equation~(\ref{dolms}) may look complicated,
but the amplification factor in square brackets on the RHS
is just a number,
as in the pretty Kaiser formula~(\ref{dkaiser}).
OK, so I lied; if you look closely,
you will see that equation~(\ref{dolms}) contains,
in the expression in the square brackets,
a function $\alpha(r)$ of radial depth $r$,
defined to be
the logarithmic derivative of the radial selection function
\begin{equation}
  \alpha(r)
  \equiv
  {\partial \ln \left[ r^2 \nbar(r) \right] \over \partial \ln r}
  \;.
\end{equation}
So the expression~(\ref{dolms}) is not as pretty as Kaiser's.
But in practice, it turns out that the $\alpha(r)$ factor gets absorbed
into another factor of $\nbar(r)$ at a previous step of the pipeline,
and so proves not to pose any special difficulty.
[If you are wondering whether equation~(\ref{dolms})
has a rigorous mathematical meaning, the answer is yes it does:
in real space,
$\alpha(r)$ is a diagonal matrix with eigenvalues $\alpha(r)$;
the symbol $\alpha(r)$ in equation~(\ref{dolms}) is
the same matrix, but expressed in $\omega lm$ space.]

\section{Quadratic Compression}
\label{quadratic}

Quadratic compression
(Tegmark 1997 \cite{T97};
Bond, Jaffe \& Knox 2000 \cite{BJK00})
is a beautiful idea
in which the information in a set of modes
is losslessly
(or almost losslessly)
compressed not all the way to cosmological parameters,
but rather to a set of power spectra.
For galaxy power spectra
the result is not one power spectrum but (at least) three:
the galaxy-galaxy, galaxy-velocity, and velocity-velocity power spectra.
I say ``at least''
because I expect that in the future
the drive to reduce systematics from luminosity bias
(more luminous galaxies are more clustered than faint galaxies)
will warrant resolution of power spectra
into ``luminous'' and ``faint'' components.

The point of reducing to power spectra rather than cosmological parameters
is that the covariance matrix $C$ in the Gaussian likelihood function
depends, by definition,
{\em linearly\/} on the prior cosmic power spectrum $p_\alpha$:
\begin{equation}
  C = \sum_\alpha C_{,\alpha} \, p_\alpha + N
  \;.
\end{equation}
Here $p_\alpha$ denotes a set of cosmic
galaxy-galaxy, galaxy-velocity, and velocity-velocity powers
at various wavenumbers,
$C_{,\alpha}$
is shorthand for the derivative $\partial C / \partial p_\alpha$,
and $N$ is the shot noise.
For example, we have typically estimated the power in 49 bins
of logarithmically-spaced wavenumbers,
so there are $49 \times 3 = 147$ power spectrum parameters $p_\alpha$
to estimate.
Normally it would be intractable to find the maximum likelihood
solution for 147 cosmological parameters,
but because the covariance $C$ depends linearly on the powers $p_\alpha$
the solution is analytic.

Although the solution is analytic,
it involves $147$ matrices $C_{,\alpha}$ each $4000 \times 4000$ in size
(for $4000$ modes),
and it still takes a bit of cunning to accomplish that solution numerically.
A good trick is to decompose the (symmetric) covariance matrix $C$ as
the product of a lower triangular matrix $L$ and its transpose $L^\transpose$
\begin{equation}
  C = L L^\transpose
  \;.
\end{equation}
The jargon name for this is Cholesky decomposition,
and there are fast ways to do it.
The Fisher matrix of the parameters $p_\alpha$ is
\begin{equation}
  F_{\alpha\beta}
  = \frac{1}{2}
    \left( L^{-1} C_{,\alpha} L^{-\transpose} \right)
    \cdot
    \left( L^{-1} C_{,\beta} L^{-\transpose} \right)
  \;.
\end{equation}

Then, from the measured amplitudes $\delta$ of the PKL modes,
form the shot-noise-subtracted {\bf quadratic estimator} $\hat q_\alpha$
\begin{equation}
\label{qalpha}
  \hat q_\alpha
  = \frac{1}{2}
  \left( L^{-1} \delta \right)^\transpose \left( L^{-1} C_{,\alpha} L^{-\transpose} \right) \left( L^{-1} \delta \right) - \hat N_\alpha
\end{equation}
in which $\hat N_\alpha$ denotes the shot noise,
the self-pair contribution to
the main term on the right hand side.
The expected mean and variance of the quadratic estimators $\hat q_\alpha$ are
\begin{equation}
\label{qbar}
  \langle \hat q_\alpha \rangle
  =
  F_{\alpha\beta} \, p_\beta
\end{equation}
\begin{equation}
  \langle \varDelta \hat q_\alpha \varDelta \hat q_\beta \rangle
  =
  F_{\alpha\beta}
  \ .
\end{equation}
Suitably scaled,
the quadratic estimates $\hat q_\alpha$
can be regarded as smoothed-but-correlated estimates of the parameters $p_\alpha$.

Given equation~(\ref{qbar}),
an estimator of power $\hat p_\alpha$,
which we call the ``raw'' estimator, can be defined by
\begin{equation}
\label{pbar}
  \hat p_\alpha
  =
  F^{-1}_{\alpha\beta} \, \hat q_\beta
\end{equation}
which is an unbiassed estimator because $\langle \hat p_\alpha \rangle = p_\alpha$.
The raw estimates are anti-correlated, with covariance
\begin{equation}
  \langle \varDelta \hat p_\alpha \varDelta \hat p_\beta \rangle
  =
  F^{-1}_{\alpha\beta}
  \ .
\end{equation}
This raw estimator $\hat p_\alpha$ exhausts the Cram\'{e}r-Rao inequality
(Paper~1, eq.~(31)),
and therefore no better estimator of power exists.

The raw estimates $\hat p_\alpha$,
along with their full covariance matrix,
contain all
the information available from the observational data,
and can be used
in a maximum likelihood analysis of cosmological parameters.
However,
if the raw estimates $\hat p_\alpha$
are plotted on a graph,
with error bars given by the square root of the
diagonal element of the covariance matrix,
$(F^{-1}_{\alpha\alpha})^{1/2}$,
then the result gives a misleadingly pessimistic impression
of the true uncertainties.
This is because in plotting errors only from the diagonal elements
of the covariance matrix,
one is effectively discarding useful information
in the cross-correlation between bins.
Thus the raw estimates of power, plotted on a graph,
appear unnecessarily pessimistic and noisy.

Whereas the quadratic estimates $\hat q_\alpha$ are correlated,
and the raw estimates $\hat p_\alpha$ is anti-correlated,
there are compromise estimators that are, like Goldilocks' porridge, just right.
These are the decorrelated estimators discussed in the next section.

It was stated above that the raw estimates $\hat p_\alpha$
and their full covariance matrix
contain all the information available from the observational data.
Actually this is not quite true,
if one uses a Gaussian approximation to the likelihood function
as a function of the parameters $p_\alpha$,
as opposed to the full likelihood function.
A principal idea behind {\bf radical compression}
(Bond, Jaffe \& Knox 2000 \cite{BJK00})
is to take functions of the parameters
arranged so as to make the likelihood function as Gaussian
in the remapped parameters as possible,
and hence to extract the last (well, almost the last) ounce of
information from the data.

\section{Decorrelation}
\label{decorrelation}

\sqrtdemofig

Decorrelation,
introduced by Hamilton (1997) \cite{H97},
and first applied (to the COBE power spectrum)
by Tegmark \& Hamilton (1998) \cite{TH98},
is another delightful concept,
yielding estimates of power at each wavenumber
that are uncorrelated with all others.
A detailed exposition is given by
Hamilton \& Tegmark (2000) \cite{HT00}.
We assume throughout this section
that the likelihood function is well-approximated as a Gaussian
in the parameters $p_\alpha$,
so that the Fisher matrix equals the inverse covariance matrix.

The idea of decorrelation applies quite generally
to any set of correlated estimates of parameters, not just to power spectra.
The left panel of
Figure~\ref{sqrtdemo}
illustrates an example of two correlated parameter estimates
$\hat p_1$ and $\hat p_2$.
The fact that the error ellipse is tilted from horizontal
indicates that the parameter estimates are correlated.

There are infinitely many linear combinations of the parameter estimates
$\hat p_1$ and $\hat p_2$
of Figure~\ref{sqrtdemo}
that are uncorrelated.
Most obviously,
the eigenvectors of the covariance matrix
-- the major and minor axes of the error ellipse --
are uncorrelated.
The decomposition of a set of parameter estimates into eigenvectors
is called {\bf Principal Component Decomposition}.

However, there are infinitely many other ways
to form uncorrelated linear combinations of correlated parameters.
The right panel of Figure~\ref{sqrtdemo}
is the same as the left panel,
but stretched out along the minor axis
so that the error ellipse becomes an error circle.
Any two orthogonal vectors on this error circle
are uncorrelated.
For example,
one possibility, shown as thick solid lines in Figure~\ref{sqrtdemo},
is to choose the two vectors on the error circle
to be parallel to the original parameter axes.
This choice of uncorrelated parameters has the merit that it is
in a sense ``closest'' to the original parameters.
When the error circle is squashed back to the orginal error ellipse
in the left panel of Figure~\ref{sqrtdemo},
the decorrelated parameters (the thick solid lines)
are no longer perpendicular, but they are nonetheless uncorrelated.
\begin{exercise}
Show that these decorrelated parameter estimates
(the ones corresponding to the thick solid lines in Figure~\ref{sqrtdemo})
are given by $F^{1/2} \hat\pvec$.
\qex
\end{exercise}

Mathematically,
decorrelating a set of correlated parameter estimates is equivalent to
decomposing their Fisher matrix as
\begin{equation}
\label{M}
  F = M^\transpose \varLambda M
\end{equation}
where $\varLambda$ is diagonal.
The matrix $M$, which need not be orthogonal, is called a {\bf decorrelation matrix}.
The parameter combinations $M \hat\pvec$ are uncorrelated beause
their covariance matrix is diagonal:
\begin{equation}
  \left\langle \varDelta (M \hat\pvec) \varDelta (M \hat\pvec)^\transpose \right\rangle
  = M \left\langle \varDelta \hat\pvec \varDelta \hat\pvec^\transpose \right\rangle M^\transpose
  = M F^{-1} M^\transpose
  = \varLambda^{-1}
  \;.
\end{equation}
Thus each row of the decorrelation matrix $M$
represents a parameter combination that is uncorrelated with all other rows.
It is possible to rescale each row of $M$
so the diagonal matrix $\varLambda$
is the unit matrix,
so that the parameter combinations $M \hat\pvec$ have unit covariance matrix.
Usually however one prefers a more physically motivated scaling.

In the case of the power spectrum,
the rows of the decorrelation matrix $M$ represent
{\bf band-power windows},
and it is sensible to normalize them to unit area
(sum of each row is one),
so that a measured power $M \hat\pvec$
can be interpreted as the power averaged over the band-power window.

Figure~\ref{sqrtdemo}
illustrates the case where the decorrelation matrix is taken to be the
{\bf square root of the Fisher matrix}
\begin{equation}
  M = F^{1/2}
  \ .
\end{equation}
Applied to the power spectrum,
this choice
(or rather, a version thereof scaled with the prior power
-- see Hamilton \& Tegmark 2000 \cite{HT00} for details)
provides nicely behaved
band-power windows that are (at least at linear scales) everywhere positive,
and concentrated narrowly about each target wavenumber,
as illustrated in Figure~\ref{sqrtf}.

By contrast,
principal component decomposition
yields broad, wiggly, non-positive band-power windows
that mix power at small and large scales
in a physically empty fashion.

\sqrtffig

\section{Disentanglement}
\label{disentanglement}

As described in \S\ref{quadratic},
quadratic compression yields estimates of not one but three power spectra,
the galaxy-galaxy (${\rm gg}$), galaxy-velocity (${\rm gv}$),
and velocity-velocity (${\rm vv}$) power spectra.
The three power spectra
are related to the true underlying matter power spectrum $P(k)$ by
(Kolatt \& Dekel 1997 \cite{KD97};
Tegmark 1998 \cite{T98};
Pen 1998 \cite{P98};
Tegmark \& Peebles 1998 \cite{TP98};
Dekel \& Lahav 1999 \cite{DL99}):
\begin{equation}
\label{Pks}
  \begin{array}{r@{\ }c@{\ }c@{\ }c@{\ }l}
    \mbox{galaxy-galaxy power} & : & P_{\rm gg}(k) & = & b(k)^2 P(k) \\
    \mbox{galaxy-velocity power} & : & P_{\rm gv}(k) & = & r(k) b(k) f P(k) \\
    \mbox{velocity-velocity power} & : & P_{\rm vv}(k) & = & f^2 P(k) \\
  \end{array}
\end{equation}
where $b(k)$ is the (possibly scale-dependent) galaxy-to-mass bias factor,
$r(k) \in [-1,1]$ is a (possibly scale-dependent)
galaxy-velocity correlation coefficient,
and $f$ is the
dimensionless linear growth rate,
which is well-approximated by
(Lahav et al 1991 \cite{LLPR91};
Hamilton 2001 \cite{H01})
\begin{equation}
  f \approx \varOmega_{\rm m}^{4/7} + (1 + \varOmega_{\rm m}/2) \, \varOmega_{\varLambda}/70
  \ .
\end{equation}
More correctly, the `velocity' here refers to minus the velocity divergence,
which in linear theory is related to the mass (not galaxy)
overdensity $\delta$ by
\begin{equation}
  f \delta + {\bf\nabla} \cdot \vvec = 0
  \;,
\end{equation}
where ${\bf\nabla}$ denotes the comoving gradient in velocity units.

The linear velocity-velocity power spectrum $P_{\rm vv}(k)$
is of particular interest because,
to the extent that galaxy flows
trace dark matter flows on large scales
(which should be true),
it offers an unbiassed measure
of the shape of the matter power spectrum $P(k)$.
Indeed if the dimensionless linear growth rate $f$ is taken as a known quantity,
then $P_{\rm vv}(k)$ provides a direct measurement
of the shape and amplitude of the matter power spectrum $P(k)$.
It should be cautioned that $P_{\rm vv}(k)$
is a direct measure of matter power only at linear scales,
where redshift distortions conform to the linear Kaiser (1987) \cite{K87} model.
At nonlinear scales,
fingers-of-god are expected to enhance the velocity-velocity power
above the predicted linear value.

Although each quadratic estimate $\hat q_\alpha$, equation~(\ref{qalpha}),
is targeted to measure a single power type and a single wavenumber
(a single parameter $p_\alpha$),
inevitably the nature of real surveys causes
each estimate $\hat q_\alpha$ to contain a mixture
of all three power spectra at many wavenumbers,
in accordance with equation~(\ref{qbar}).

What one would like to do is to disentangle the three power spectra,
projecting out an unmixed version of each.
The problem is somewhat similar to the CMB problem
of forming disentangled
$TT$, $TE$, $EE$, and $BB$ power spectra
from the amplitudes of fluctuations
in temperature ($T$),
and in electric ($E$) and magnetic ($B$) polarization modes
($TB$ and $EB$ cross power spectra are expected to vanish because
$B$ modes have opposite parity to $T$ and $E$ modes).

Now the raw estimates of power $\hat p_\alpha$,
equation~(\ref{pbar}),
are already disentangled,
and as remarked in \S\ref{quadratic},
they and their covariance matrix contain (almost) all the information
in the observational data.
Thus if the aim is merely to do a cosmological parameter analysis,
then it is fine to stop at the raw powers $\hat p_\alpha$.

However, before leaping to cosmological parameters,
it is wise to plot the
${\rm gg}$, ${\rm gv}$, and ${\rm vv}$
power spectra explicitly.

One possibility would be to plot each power spectrum
marginalized over the other two.
The drawback with this is that
marginalization effectively means discarding information
contained in the correlations between the power spectra,
and discarding this information leads to unnecessarily
noisy estimates of power.

Instead, we have adopted the following procedure.
First,
we decorrelate the entire set of estimates of power
(three power spectra at each of many wavenumbers)
with the square root of the Fisher matrix
(prescaled with the prior power
-- see Hamilton \& Tegmark 2000 \cite{HT00}).
The result is three power spectra
every point of which is uncorrelated with every other
(with respect to both type and wavenumber).
The three power spectra are uncorrelated, but not disentangled:
each uncorrelated power spectrum contains contributions from all three types,
${\rm gg}$, ${\rm gv}$, and ${\rm vv}$.
To disentangle them,
we multiply the three uncorrelated power spectra at each wavenumber
with a matrix that, if the prior is correct,
yields pure ${\rm gg}$, ${\rm gv}$, and ${\rm vv}$ power spectra.
The resulting band-power spectra and their error bars
represent the information in the survey about as fairly as can be.
The three power spectra at each wavenumber are correlated,
but uncorrelated with the power spectra at all other wavenumbers.

Figure~\ref{sqrtf}
illustrates the resulting band-power windows
for each of the
${\rm gg}$, ${\rm gv}$, and ${\rm vv}$
power spectra in the PSCz survey.
The ${\rm gg}$ band-power window,
for example, contains contributions from ${\rm gv}$, and ${\rm vv}$ powers,
but those contributions should cancel out if the prior power is correct.
Note that cancellation invokes the prior only weakly:
the off-type contributions will cancel
provided only that the true power has the same shape
(not the same normalization)
as the prior power over a narrow range
(because the band-power windows are narrow).


\xikfullfig

Figure~\ref{xikfull}
shows the galaxy-galaxy, galaxy-velocity, and velocity-velocity power spectra
$P_{\rm gg}$,
$P_{\rm gv}$,
and
$P_{\rm vv}$
measured in this fashion from the PSCz survey
(the galaxy-galaxy power spectrum $P_{\rm gg}$ shown here
appears in 
Hamilton, Tegmark \& Padmanabhan 2000 \protect\cite{HTP00},
but $P_{\rm gv}$ and $P_{\rm vv}$
have not appeared elsewhere).
As can be seen,
the galaxy-velocity and velocity-velocity power spectra
are noisy, but well detected.
Nonlinear fingers-of-god
cause the galaxy-velocity power $P_{\rm gv}$
to go negative at $k \ga 0.2 \, h \, {\rm Mpc}^{-1}$,
and at the same time enhance
the velocity-velocity power $P_{\rm vv}$ somewhat.

%

\section{Conclusion}
\label{conclusion}

In its simplest form,
measuring the galaxy power spectrum is almost trivial:
bung the survey in a box,
Fourier transform it,
and measure the resulting power spectrum
(Baumgart \& Fry 1991) \cite{BF91}.
A variant of this method,
the Feldman, Kaiser \& Peacock (1994) \cite{FKP94} method,
in which the galaxy density is weighted
by a weighting which is near-minimum-variance
subject to some not-necessarily-true assumptions
before Fourier transforming it,
offers a fast and not-so-bad way to measure power spectra,
good for when maximum likelihood would be overkill.

However, the ``right'' way to measure power spectra,
at least at large, linear scales,
is to use Bayesian maximum likelihood analysis.
Maximum likelihood methods for measuring galaxy power spectra
were pioneered by
Fisher, Scharf \& Lahav (1994) \cite{KSL94}
and Heavens \& Taylor (1995) \cite{HT95},
and have seen considerable refinement in recent years.
Maximum likelihood methods require a lot more work than traditional methods.
The pay-off is not necessarily more precision
(the error bars are all too often larger
than claimed error bars from more primitive methods),
but more precision in the precision.
That is,
you can have some confidence that the derived uncertainties
reliably reflect the information in the data,
no more and no less.
This is essential when measurements
are to be used in estimating cosmological parameters.
Moreover maximum likelihood methods provide a general way to deal
with all the complications of real galaxy surveys,
as spherical redshift distortions, Local Group flow,
variable extinction, and such-like.

\section*{Acknowledgements}
This work was supported in part by
NASA ATP award NAG5-10763 and by NSF grant AST-0205981,
and of course by this wonderful summer school.

\end{document}